\documentclass[eqsecnum,showpacs,showkeys,nofootinbib,aps]{revtex4}
\renewcommand{\theequation}{\arabic{equation}}
\def\beq{\begin{equation}}
\def\eeq{\end{equation}}
\def\bea{\begin{eqnarray}}
\def\eea{\end{eqnarray}}
\def\nn{\nonumber}

\begin{document}
\title{Kaluza-Klein mass spectra on extended dimensional branes}
\author{Soon-Tae Hong}
\email{soonhong@ewha.ac.kr} 
\affiliation{Department of Science Education and Research Institute for Basic Sciences,\\
Ewha Womans University, Seoul 120-750 Korea}
\date{\today}%

\begin{abstract}
We study the hierarchy problem on the four- and five-branes which are 
constructed by attaching a circle and a sphere to the standard three-brane, 
respectively. The effective masses in their excited spectra on these extended dimensional branes imply 
intriguing characteristics associated with the quantization of the compact circular 
and spherical manifolds. In particular, their lightest effective
masses are shown to suppress exponentially with respective to the Planck mass, similar to the
three-brane case. We also investigate the effective couplings in these extended dimensional branes.
\end{abstract}
\maketitle

\section{introdution}
\setcounter{equation}{0}
\renewcommand{\theequation}{\arabic{section}.\arabic{equation}}

There have been lots of progresses and discussions on the
Randall-Sundrum (RS) model~\cite{RS1,RS2}. The RS model on the three-brane has been analyzed in
terms of the Kaluza-Klein (KK) decompositions~\cite{kaluza,klein} to yield the effective
scale of couplings~\cite{gold1,gold2}. On the RS brane world, the
population evolution and primordial black holes have been
investigated to study a modified evaporation law of the black
hole~\cite{guedens02}. The string excitations in the RS effective
model for electroweak symmetry breaking also has been studied in
the string theory to investigate the mass of the string states
associated with the number of colors. Here it has been shown that
there exist strong bounds on the mass of new string states to
yield the collider signals~\cite{reece10}. Recently, the RS model
has been revisited in which the hierarchy between the Planck and
electroweak scales, and the Higgs field confined to a brane at the
infrared scale are well defined. Specifically, it has been
suggested that in the Standard Model confined to the TeV brane,
the KK modes of the graviton in the RS scenario could be
detectable in the Large Hadron Collider (LHC)  detector, if the TeV
brane energy scale is sufficiently low~\cite{kelly11}.

On the other hand, the string singularities, which are the string
theory~\cite{witten87,pol98} version of the Hawking-Penrose
singularities~\cite{hawking}, have been applied to the early
universe with an arbitrary higher
dimensionality~\cite{hong07,hong08,hong11}. In this higher
dimensional stringy cosmology, the expansion of the universe has
been explained by exploiting Hawking-Penrose type singularity in
geodesic surface congruences for the timelike and null
strings~\cite{hong07,hong08}. Next, the twist and shear have been
studied in terms of the expansion of the universe. Moreover, as
the early universe evolves with expansion rate, the twist of the
stringy congruence decreases exponentially and the initial twist
value should be large enough to sustain the rotations of the
ensuing universe, while the effects of the shear are negligible to
produce the isotropic and homogeneous universe~\cite{hong11}. By
exploiting the phantom field, the evolution of cosmomogy has been
also studied in higher dimensional spacetime~\cite{hong08c}.

Recently, the RS model has been investigated on the extended dimensional spacetime 
associated with $AdS_{5}\times M^{d}$ where $d$ is the extended dimensionality 
of the branes. Specifically, the cases of $d=1$ and 
$d=2$ have been studied in~\cite{rizzo03,rizzo08}. 
In the $d$-dimensional flat manifold, they have accommodated localized 
fermions to investigate proton decay and flavor problems~\cite{rizzo03}. 
Exploiting the four-brane scheme, which is constructed by attaching the additional 
$S^{1}$ to the standard RS model, the LHC phenomenology has been analyzed in~\cite{rizzo03,rizzo08}. 

The motivation of this work is to investigate the 
natures, such as mass spectra corresponding to excitations along $S^{1}$ and $S^{2}$, 
of the extended dimensional spacetime related with the RS model as in~\cite{rizzo03,rizzo08}.
We emphasize that in our work we will formulate explicitly the effective masses by following 
the KK decomposition scheme. Moreover, we construct the effective couplings in six- and seven-dimensions. 

Quite recently, the RS model has been also investigated in the framework of the three-brane embedded in 
six dimensional multiply warped spacetime associated with two orbifold compact directions~\cite{koley10}. 
In this scenario, the effective scalar self-couplings of 
all orders have been shown to be enhanced since this model have an extra warped dimension. 
Moreover, this enhancement of the effective couplings have been presented to make the multiply warped model 
phenomenologically more attractive. They also have evaluated 
the KK mode masses of  a bulk scalar field in a four-brane world with two warped extra dimensions.

In this paper we will generalize the KK modes in the RS model with two 
three-branes~\cite{gold1,gold2} to the cases of two four-branes and five-branes, 
to construct the corresponding effective mass spectra of the reduced three-branes. 
To do this, we will exploit the novel aspects of bulk fields in the KK decompositions in the RS
model, and we will next investigate the hierarchy problems
on these extended dimensional branes, together with considering the effective couplings in six- and seven-dimensions.\\

\section{Four-brane cosmology and effective mass spectrum}
\setcounter{equation}{0}
\renewcommand{\theequation}{\arabic{section}.\arabic{equation}}

In the standard RS model~\cite{RS1,RS2,gold1,gold2}, they investigated some interesting aspects
originated from the extra dimension to yield successful theoretical predictions. In fact,
they used one more extra dimensional model, and in their scenarios there exist the three-brane
cosmology connected by this one dimensional manifold. We now extend the dimensionality
of the brane to the next larger space, namely the four-brane case. On the other hand, the ordinary (4+2)-dimensional
cosmology has been shown to possess some intriguing aspects in the stringy early universe~\cite{hong08}.
In the limit of the four-brane case in which the total manifold has $D=6$ dimensions, we observed that
the strong energy condition in the stringy cosmology
\beq
\frac{D-4}{D-2}\rho+\frac{D}{D-2}P\ge 0,
\eeq
becomes the first one of the point particle strong energy conditions
$\rho+3P\ge 0$ and $\rho+P\ge 0$ for the massive particle~\cite{hong08}. In that sense, the four-brane cosmology may reveal
a critical feature and leads us an interesting physics.

In order to investigate the bulk fields in the $(4+2)$-dimensional RS model, we
introduce the metric of the specific form~\cite{rizzo03}
\beq
ds_{4+2}^{2}= e^{-2 k r_{c} |\alpha|}(\eta_{\mu\nu} dx^{\mu} dx^{\nu}-R^{2}d\phi^{2})
- r_{c}^{2} d\alpha^{2}. \label{metric4}
\eeq
Here $x^{\mu}$ ($\mu=0,1,2,3$) are the coordinates on the $(3+1)$-dimensional  Lorentzian spacetime ${\cal M}$. 
We have the $S^{1}(\alpha)/{\bf Z}_{2}$ orbifold constraint to yield $(x^{\mu},\phi,\alpha)
=(x^{\mu},\phi,-\alpha)$. The four-branes are imposed to be located at $\alpha=0$ and $\alpha=\pi$.
The $r_{c}$ and $k$ are the the size of the extra dimension in the $\alpha$-direction
and the order of Planck scale. The $\phi$ is the coordinate defined on the compact manifold 
$S^{1}(\phi)$ of radius $R$ and $0\le \phi\le 2\pi$. We then have an
action for a free scalar field in the bulk as follows
\beq
S=\frac{1}{2}\int_{\cal M} d^{4}x~\int_{0}^{2\pi}d\phi~\int_{-\pi}^{\pi} d\alpha~\sqrt{|g|}
\left(g^{AB}\partial_{A}\Psi^{*}\partial_{B}\Psi - m^{2}|\Psi|^{2}\right),
\label{bulkaction4}
\eeq
where $g_{AB}$ ($A,B=\mu,\phi,\alpha$) are given by 
\beq
g_{AB}={\rm diag}~(e^{-2kr_{c}|\alpha|}\eta_{\mu\nu},-e^{-2kr_{c}|\alpha|}R^{2}, -r_{c}^{2}),
\eeq 
from which we obtain $\sqrt{|g|}=e^{-5\sigma}Rr_{c}$ with $\sigma=kr_{c}|\alpha|$. After some algebra, 
we arrive at
\bea
S&=&\frac{1}{2}\int_{\cal M} d^{4} x~\int_{0}^{2\pi}d\phi~\int_{-\pi}^{\pi} d\alpha~Rr_{c}
\left(e^{-3\sigma}\eta^{\mu\nu}\partial_{\mu}\Psi^{*}
\partial_{\nu}\Psi\right.
\nn\\
&&
\left.-\frac{e^{-3\sigma}}{R^{2}}\partial_{\phi}\Psi^{*}\partial_{\phi}\Psi
+\frac{1}{r_{c}^{2}}\Psi^{*}\partial_{\alpha}\left(e^{-5\sigma}\partial_{\alpha}
\Psi\right)-m^2 e^{-5\sigma}|\Psi|^{2}\right).\nn\\
\label{action5}\eea

Next, we consider the KK decompositions of $\Psi(x,\phi,\alpha)$ along the compact $\alpha$ manifold
in this four-brane case by introducing
\bea
\Psi(x,\phi,\alpha)&=&\sum_{n}\psi_{n}(x,\phi)\frac{y_{n}(\alpha)}{\sqrt{r_{c}}},\nn\\
\Psi^{*}(x,\phi,\alpha)&=&\sum_{n}\psi_{n}^{*}(x,\phi)\frac{y_{n}(\alpha)}{\sqrt{r_{c}}},~~~(n=1,2,3,...),
\label{psixphi4}
\eea
where the fields $y_{n}(\alpha)$ satisfy the normalization condition
\beq
\int_{-\pi}^\pi d\alpha~e^{-3\sigma} y_{n}(\alpha)
y_{m}(\alpha)=\delta_{nm}
\eeq
and the differential equation related to the Bessel functions
\beq
-\frac{1}{r_{c}^{2}}\frac{d}{d\alpha}\left(e^{-5\sigma}
\frac{dy_n}{d\alpha}\right)+m^{2}e^{-5\sigma}y_{n}
= m_{n}^{2}e^{-3\sigma} y_{n}.
\label{eigen}
\eeq
Here one notes that the mass $m_{n}$ is introduced as eigenvalue corresponding to the eigenfunction
$y_{n}$ associated with the differential equation in (\ref{eigen}). Introducing new variables
\beq
z_{n}=\frac{m_{n}}{k}e^{\sigma},~~~f_{n} =e^{-5\sigma /2}y_{n},
\eeq 
we rewrite (\ref{eigen}) for
$\alpha\neq 0, \pm\pi$ as
\begin{equation}
z_{n}^{2}\frac{d^{2}f_{n}}{d z_{n}^{2}} + z_{n} \frac{d f_{n}}{dz_{n}} +
\left(z_{n}^{2}
-\left(\frac{5}{2}\right)^{2}-\frac{m^{2}}{{k}^2}\right)f_{n}=0,
\label{zfz}
\end{equation}
whose solutions are given by the Bessel functions
\beq 
f_{n}(z_{n})=\frac{1}{N_{n}}\left(J_{\nu} \left(z_{n}\right) + b_{n\nu}
Y_{\nu} \left(z_{n}\right)\right),
\eeq 
where $N_{n}$ is a normalization factor to be fixed as follows, 
\bea
N_{n}&=&\frac{e^{kr_{c}\pi}}{\sqrt{kr_{c}}}~A_{n},\nn\\
A_{n}&=&J_{\nu}(x_{n\nu})\left(1+\frac{(5/2)^{2}-\nu^{2}}{x_{n\nu}^{2}}\right),
\label{an}
\eea
and $J_{\nu}$ and $Y_{\nu}$
are the first- and second-kind Bessel functions of order
\beq
\nu=\left(\left(\frac{5}{2}\right)^{2}+{m^{2}\over
k^{2}}\right)^{1/2}.
\eeq

The solutions $y_{n}$ of the Bessel function 
(\ref{eigen}) are given by 
\beq
y_{n}=\frac{e^{5\sigma/2}}{N_{n}}\left(J_{\nu}\left(\frac{m_{n}}{k} e^{\sigma}\right) 
+b_{n\nu}Y_{\nu}\left(\frac{m_{n}}{k} e^{\sigma}\right)\right). \label{yn} 
\eeq
Now we exploit the condition that $\frac{dy_{n}}{d\alpha}$ is continuous at the orbifold fixed points 
$(\alpha=0,\pm\pi)$ as in~\cite{gold1}. In particular, at $\alpha=0$ this continuity condition 
produces $b_{n\nu}$ as follows 
\beq
b_{n\nu}=-\frac{\frac{5}{2}J_{\nu}\left(\frac{m_{n}}{k}\right)
+\frac{m_{n}}{k}J_{\nu}^{\prime}\left(\frac{m_{n}}{k}\right)}
{\frac{5}{2}Y_{\nu}\left(\frac{m_{n}}{k}\right)
+\frac{m_{n}}{k}Y_{\nu}^{\prime}\left(\frac{m_{n}}{k}\right)},
\eeq 
where 
\beq
J_{\nu}^{\prime}(x)=\frac{dJ_{\nu}^{\prime}(x)}{dx},~~~
Y_{\nu}^{\prime}(x)=\frac{dY_{\nu}^{\prime}(x)}{dx}.
\eeq 
Next, another continuity condition at $\alpha=\pi$ 
yields the analysis that $\frac{m_{n=1}}{k}e^{kr_{c}\pi}$ and $\frac{m}{k}$ are of order unity, similar to the 
standard RS model case~\cite{gold1}. We then obtain that $m$ defined in the six-dimensional spacetime and 
the lightest mass $m_{n=1}$ of the KK modes in the five-dimensional spacetime are related to each other as follows
\beq
m_{n=1}\cong me^{-kr_{c}\pi}.
\label{mnme}
\eeq
Assuming that $m$ is of order of the Planck scale mass $M_{Planck}$, and choosing the value 
$kr_{c}\cong 12$ as in~\cite{gold1}, one can readily see that $m_{n=1}$ is of order of 1 TeV.
 
We proceed to arrive at the action of the form
\beq
S=\frac{1}{2}\sum_{n} \int_{\cal M} d^{4}x~\int_{0}^{2\pi}d\phi~R
\left(\eta^{\mu\nu}\partial_{\mu} \psi_{n}^{*} \partial_{\nu} \psi_{n}
-\frac{1}{R^{2}}(\partial_{\phi}\psi_{n}^{*}\partial_{\phi}\psi_{n})^{2}- m_{n}^{2} |\psi_{n}|^{2}\right).
\label{action4}
\eeq
We again perform further the KK decompositions of $\psi_{n}(x,\phi)$ along the $\phi$ direction
by exploiting the following ansatz for a given $n$ 
\bea
\psi_{n}(x,\phi)&=&\sum_{p}X_{np}(x)\frac{e^{ip\phi}}{\sqrt{2\pi R}},\nn\\
\psi_{n}^{*}(x,\phi)&=&\sum_{p}X_{np}(x)\frac{e^{-ip\phi}}{\sqrt{2\pi
R}},~~~(p=0,\pm 1,\pm 2,...). \label{psinx} 
\eea 
The action
(\ref{action4}) is then rewritten as \beq S=\frac{1}{2}\sum_{n,p}
\int_{\cal M} d^{4}x~
\left(\eta^{\mu\nu}\partial_{\mu}X_{np}\partial_{\nu}X_{np} -
m_{np}^{2}X_{np}^{2}\right), \label{action42} \eeq where the
effective mass $m_{np}$ on the reduced three-brane is given by, with 
$n=1,2,3,...$ and $p=0,\pm 1,\pm 2,...$ 
\beq
m_{np}=\left(m_{n}^{2}+\frac{p^{2}}{R^{2}}\right)^{1/2}.
\label{mnp} \eeq 
It is interesting to note that, in the large $R$
limit with $R$ being the radius of the extended dimension of the
four-brane, $m_{np}$ on the three-brane reduces to the effective mass
$m_{n}$ on the four-brane. Assuming again that $m$ is of order of the
Planck scale mass $M_{Planck}$, from (\ref{mnme}) and (\ref{mnp})
we observe that the mass $m_{n=1,p=0}$ on the reduced three-brane is
suppressed with respect to the $M_{Planck}$ as follows 
\beq
m_{n=1,p=0}\cong M_{Planck}e^{-kr_{c}\pi},
\label{mnp11} \eeq 
which is also of order of 1 TeV with the choice of $kr_{c}\cong 12$ as in~\cite{gold1}, 
and whose characteristics will be discussed later, together with those of the five-brane case.\\

\section{Five-brane and Kaluza-Klein decompositions}
\label{antiparticle}
\setcounter{equation}{0}
\renewcommand{\theequation}{\arabic{section}.\arabic{equation}}

Next, we consider the five-brane cosmology whose metric is given by~\cite{rizzo03}
\beq
ds_{5+2}^{2}= e^{-2 k r_{c} |\alpha|}(\eta_{\mu\nu} dx^{\mu} dx^{\nu}-R^{2}(d\theta^{2}
+\sin^{2}\theta d\phi^{2}))
- r_{c}^{2} d\alpha^2 \label{metric5}
\eeq
to yield $\sqrt{|g|}=e^{-6\sigma}R\sin\theta~r_{c}$. On the five-brane, the action (\ref{bulkaction4}) 
for the four-brane case is now modified as~\footnote{Even though in (\ref{bulkaction5}) we have used
the same notation $m$ for the mass defined in the seven-dimensional action as
 that for the mass in the six-dimensional action in (\ref{bulkaction4}), the differences 
between the masses in $D=6$ and $D=7$ models are understood in the context.}
\beq
S=\frac{1}{2}\int_{\cal M} d^{4}x~\int_{0}^{\pi}d\theta~\int_{0}^{2\pi}d\phi~\int_{-\pi}^{\pi} d\alpha~\sqrt{|g|}
\left(g^{AB}\partial_A \Psi \partial_B \Psi - m^2 \Psi^{2}\right).
\label{bulkaction5}
\eeq
After some algebra, the above action has the following form
\bea S&=&\frac{1}{2}\int_{\cal M} d^{4} x~\int_{0}^{\pi}d\theta~\int_{0}^{2\pi}d\phi~\int_{-\pi}^{\pi}
d\alpha~R^{2}\sin\theta~r_{c}\left(e^{-4\sigma}\eta^{\mu\nu}\partial_{\mu}\Psi\partial_{\nu}\Psi
-\frac{e^{-4\sigma}}{R^{2}}(\partial_{\theta}\Psi)^{2}
\right.\nn\\
&&\left.-\frac{e^{-4\sigma}}{R^{2}\sin^{2}\theta}(\partial_{\phi}\Psi)^{2}
+\frac{1}{r_{c}^{2}}\Psi\partial_{\alpha}\left(e^{-6\sigma}\partial_{\alpha}
\Psi\right)-m^2 e^{-6\sigma}\Psi^2\right). 
\label{action52}\eea

Now we introduce KK modes along the $\alpha$ and $\phi$ directions by exploiting
\beq
\Psi(x,\theta,\phi,\alpha)=\sum_{n,p}X_{np}(x,\theta)\frac{e^{ip\phi}y_n(\alpha)}{\sqrt{2\pi Rr_c}},
\eeq
with $n=1,2,3,...$ and $p=0,\pm 1,\pm 2,...$. Here the fields $y_{n}(\alpha)$ defined in the 
five-brane scenario now fulfill the condition
\beq
\int_{-\pi}^\pi d\alpha~e^{-4\sigma} y_{n}(\alpha)
y_{m}(\alpha)=\delta_{nm}
\eeq and the differential equation
\beq
-\frac{1}{r_{c}^{2}}\frac{d}{d\alpha}\left(e^{-6\sigma}
\frac{dy_n}{d\alpha}\right)+m^{2}e^{-6\sigma}y_{n}
= m_{n}^{2}e^{-4\sigma} y_{n}.
\label{eigen2}
\eeq
Equation (\ref{eigen2}) describes the relation between the matter mass $m$ defined on the 
$D=7$ total manifold with the additional two dimensions (with respect to the standard RS model) and 
the mass $m_{n}$ corresponding to the eigenvalue associated with the 
eigenfunction $y_{n}$ in (\ref{eigen2}). Here, one notes that, as in the 
case of the four-brane scenario, we again obtain 
the effective mass suppression of the form (\ref{mnme}). We also observe that (\ref{eigen2}) is 
different from (\ref{eigen}) due to the increased dimensionality.

We then arrive at
\bea
S&=&\frac{1}{2}\sum_{n,p} \int_{\cal M} d^{4}x~\int_{0}^{\pi}d\theta~R\sin\theta
\left(\eta^{\mu\nu}\partial_{\mu}X_{np}\partial_{\nu}X_{np}\right.\nn\\
&&\left.-\frac{1}{R^{2}}(\partial_{\theta}X_{np})^{2}
-\frac{p^{2}}{R^{2}\sin^{2}\theta}X_{np}^{2}-m_{n}^{2}X_{np}^{2}\right).
\label{action53}
\eea
We again perform the KK decompositions along the compact $\theta$ coordinate with the following ansatz
\beq
X_{np}(x,\theta)=\sum_{l}R_{npl}(x)\frac{N_{lp}}{\sqrt{R}}P_{l}^{p}(\theta),~~~(l=0,1,2,3,...)
\label{psixphi52}
\eeq
where $P_{l}^{p}(\theta)$ are the associated Legendre functions satisfying the identities
\bea
\frac{1}{\sin\theta}\frac{d}{d\theta}\left(\sin\theta\frac{dP_{l}^{p}}{d\theta}\right)
-\frac{p^{2}}{\sin^{2}\theta}P_{l}^{p}&=&-l(l+1)P_{l}^{p},\nn\\
\int_{0}^{\pi}d\theta~\sin\theta~P_{l}^{p}(\theta)P_{l^{\prime}}^{p}(\theta)&=&N_{lp}^{-2}\delta_{ll^{\prime}},
\eea 
with 
\beq
N_{lp}=\left(\frac{2l+1}{2}\cdot\frac{(l-p)!}{(l+p)!}\right)^{1/2}.
\eeq 
Here, for any given $l$, there are $(2l+1)$ possible values of $p$: $p=-l, -l+1,..., -1, 0, 1,..., l-1, l$.

Using the above identities related with the associated Legendre functions, 
we then rewrite the action (\ref{action53}) in the form
\beq
S=\frac{1}{2}\sum_{n,p,l} \int_{\cal M} d^{4}x~\left(\eta^{\mu\nu}\partial_{\mu}R_{npl}\partial_{\nu}R_{npl}
-m_{nl}^{2}R_{npl}^{2}\right)
\eeq 
where the effective mass spectrum on the reduced three-brane is given by, with $n=1,2,3,...$ and $l=0,1,2,...$
\beq
m_{nl}=\left(m_{n}^{2}+\frac{l(l+1)}{R^{2}}\right)^{1/2}.
\label{mnl}
\eeq
It is interesting to see that, even though we have enlarged the dimensionality by two with respect to
the three-brane RS model, the effective mass spectrum in (\ref{mnl}) in the five-brane case at hand
has the form similar to the four-brane result (\ref{mnp}). This phenomena originate from the fact that the
harmonic spectrum of $S^{2}$ manifold is described in term of only one quantum number $l$ $(l=0,1,2,3,...)$ 
in (\ref{mnl}). Following the algorithms given above, from (\ref{mnme}) and (\ref{mnl}) 
we obtain that the mass $m_{n=1,l=0}$ on the reduced three-brane is suppressed exponentially again with respect to the 
Planck mass $M_{Planck}$
\beq
m_{n=1,l=0}\cong M_{Planck}e^{-kr_{c}\pi},
\label{mnl10}
\eeq
which is the same as the three-brane RS model result~\cite{gold1}.\\

\section{Effective scales of couplings}
\label{antiparticle}
\setcounter{equation}{0}
\renewcommand{\theequation}{\arabic{section}.\arabic{equation}}

Now we consider the self interaction whose action in the six dimensions is 
given by
\beq
S_{\rm int}=\int_{\cal M} d^{4}x~\int_{0}^{2\pi}d\phi~\int_{-\pi}^{\pi} d\alpha~\sqrt{|g|}
\frac{\lambda}{M^{4m-6}}|\Psi|^{2m},
\label{intaction6}
\eeq
where we have introduced the power $m$ in order to guarantee the dimensionless of the above action. Here the 
action (\ref{intaction6}) and the action (\ref{intaction7}) below are generalized ones of 
the three-brane RS model action proposed in~\cite{gold1}. Using the KK decompositions (\ref{psixphi4}) 
and (\ref{psinx}) of $\Psi(x,\phi,\alpha)$ along $\alpha$ and $\phi$ manifolds in the four-brane case, we arrive at
\bea
S_{\rm int}&=&\int_{\cal M} d^{4}x~\int_{0}^{2\pi}d\phi~\int_{-\pi}^{\pi} d\alpha~e^{-5\sigma}R r_{c}~
\frac{\lambda}{M^{4m-6}}\nn\\
&&\cdot\left(X_{np}(x)\frac{e^{-ip\theta}}{\sqrt{2\pi R}}\frac{y_{n}(\alpha)}{\sqrt{r_{c}}}\right)^{m}
\left(X_{np}(x)\frac{e^{ip\theta}}{\sqrt{2\pi R}}\frac{y_{n}(\alpha)}{\sqrt{r_{c}}}\right)^{m}.
\label{intaction62}
\eea
Here we do not sum over the indices $n$ and $p$ since we consider the lightest KK states only. 
Exploiting (\ref{intaction62}), we define the effective coupling constant $\lambda_{\rm eff}$ as follows
\beq
S_{\rm int}=\int_{\cal M} d^{4}x\lambda_{\rm eff}~X_{np}^{2m}
\label{intaction63}
\eeq 
where 
\beq
\lambda_{\rm eff}=\int_{0}^{2\pi}d\phi~\int_{-\pi}^{\pi}d\alpha~e^{-5\sigma}R r_{c}~\frac{\lambda}{M^{4m-6}}
\left(\frac{1}{\sqrt{2\pi R}}\frac{y_{n}(\alpha)}{\sqrt{r_{c}}}\right)^{2m}.
\eeq

After some algebra, we obtain the effective coupling constant of the form
\beq
\lambda_{\rm eff}=2\lambda\left(\frac{k}{2\pi RM}\right)^{m-1} \left(Me^{-kr_{c}\pi}\right)^{-3m+5}
\int_{0}^{1} dr~r^{5m-6}\left(\frac{J_{\nu}(x_{n\nu}r)}{A_{n}}\right)^{2m}
\label{leffective6}
\eeq
where we have used the identity (\ref{an}).

Next, we take the five-brane case into consideration to investigate the effective coupling constant. To do this, 
we start with the action
\beq
S_{\rm int}=\int_{\cal M} d^{4}x~\int_{0}^{\pi}d\theta~\int_{0}^{2\pi}d\phi~\int_{-\pi}^{\pi} d\alpha~\sqrt{|g|}~
\frac{\lambda}{M^{5m-7}}|\Psi|^{2m}.
\label{intaction7}
\eeq
Exploiting the definition of $\lambda_{\rm eff}$ in (\ref{intaction63}), after algebraic manipulations we are left with
\beq
\lambda_{\rm eff}=8\lambda\left(\frac{k}{\pi RM}\right)^{m-1} \left(Me^{-kr_{c}\pi}\right)^{-4m+6}
\int_{0}^{1} dr~r^{6m-7}\left(\frac{J_{\nu}(x_{n\nu}r)}{A_{n}}\right)^{2m}.
\label{leffective7}
\eeq
Here we emphasize that the masses $M$ in the actions (\ref{intaction6}) and (\ref{intaction7}) are depressed 
by the factor $e^{-kr_{c}\pi}$ to yield $Me^{-kr_{c}\pi}$ in the four dimensional effective 
coupling constants $\lambda_{eff}$ in (\ref{leffective6}) and (\ref{leffective7}), respectively. These results are consistent 
with the standard three brane RS model result in \cite{gold1}. Moreover, these effective couplings go well with 
the results (\ref{mnp11}) and (\ref{mnl10}) in the previous sections.\\

\section{Conclusions and discussions}
\label{antiparticle}
\setcounter{equation}{0}
\renewcommand{\theequation}{\arabic{section}.\arabic{equation}}

Now, we have comments to address. In (\ref{mnp}) and (\ref{mnl}), 
we recognize that the effective masses of the excited spectrum of the four-brane and five-brane 
depend on the radii of the additional circular and spherical manifolds included in their branes.
These characteristics originate from the quantizations of the compact circular and spherical spaces 
in the four-brane and five-brane, respectively. Moreover, it is intriguing to observe that 
the $R$ dependence in (\ref{mnp}) and (\ref{mnl}) implies that, as $R$ gets smaller, its suppression 
effect on the reduced three-brane masses becomes weaker. 

More specifically, we consider the excited states in (\ref{mnp}) and (\ref{mnl}) in two extreme limits of the radius $R$. 
First, in the limit that $R$ goes to infinity, the effective masses $m_{np}$ and $m_{nl}$ become those in 
(\ref{mnp11}) and (\ref{mnl10}), which are the same as the result of the standard RS model KK modes~\cite{gold1}.
These aspects seems to imply that the hierarchy problem reduces to the problem of expounding the reason for the 
large $R$ value~\cite{arkani98,rubakov01}. However, we observe that, in the standard RS model KK modes~\cite{gold1} which 
is a subset of our two models, the hierarchy problem can be explained without introducing the extended compact 
manifolds of the circle or the sphere possessing the large radus $R$ attached to the standard five-dimensional RS model. Second, in the limit that $R$ approaches to zero, the effective masses $m_{np}$ and $m_{nl}$ become infinity 
to destroy $m_{np}$ and $m_{nl}$ formulas in (\ref{mnp}) and (\ref{mnl}). These two extreme 
features of the four-brane and five-brane scenarios do not appear in the standard RS model KK mode analysis
for the three-brane in~\cite{gold1} to yield the interesting characteristics of the four- and five-branes. 
One of our main goals is to demonstrate these new features. Here one notes that the extended spaces attached 
to the standard five-dimensional RS model are not those of the KK type compactification 
scheme~\cite{kaluza,klein} which has hidden dimensions with vanishing compact manifold radius limit. 

Moreover, the results in (\ref{mnp}) and (\ref{mnl}) with $R$ residing in the interval between the above two extreme 
limits  show that, in order for $m_{np}$ and $m_{nl}$ to be consistent with the standard RS model, 
$R$ should be of order of $pM_{Planck}^{-1}e^{kr_{c}\pi}$ and $l(l+1)M_{Planck}^{-1}e^{kr_{c}\pi}$ for the given excited 
states (with the quantum numbers $p$ and $l$) of the four- and five-branes, respectively. We also 
observe that, in the case of the first excited state with $n=1$ and $p=\pm 1$ in (\ref{mnp}) for instance, we have 
$m_{n=1,p=\pm 1}\cong\left(m^{2}e^{-2kr_{c}\pi}+\frac{1}{R^{2}}\right)^{1/2}$ which is also lighter than the 
bare multidimensional mass $m$ due to the warp factor $e^{-kr_{c}\pi}$ as far as the second term is smaller than 
the first one. Here, one notes that all the above nontrivial results associated with the topology of the 
two-sphere for instance cannot be explained in the previous 
works~\cite{rubakov01,rubakov00} where they consider the metric of the straightforward generalization of (\ref{metric4}).
Finally we reemphasize that, in the specific cases of $p=0$ and $l=0$, even though we still have the nonvanishing $R$ in the 
formulas (\ref{mnp}) and (\ref{mnl}), the $m_{np}$ and $m_{nl}$ reduce to the standard RS model result which is 
the same as (\ref{mnp11}) and (\ref{mnl10}).

Next, we have comments to address. First, we note in (\ref{mnp}) and (\ref{mnl}) with $p\neq 0$ and $l\neq 0$ that as far as $R$ goes to zero, the curvatures associated with $R$ become bigger since the curvatures are described as $1/R^{2}$. 
This curvature is similar to (different from) the scalar curvature in general relativity in the case of $S^{2}$ ($S^{1}$) 
extra dimension. Second, for the cases of $R=0$, the attached compact  manifolds $S^{1}$ and $S^{2}$ disappear so that the four-brane and five-brane can reduce to the standard three-brane. Third, as $R$ goes to infinity, the curvatures for both cases of $S^{1}$ and $S^{2}$ extra dimensions become smaller as expected. Fourth, for the cases of $p=0$ and $l=0$, the ground states dictated by the quantization yield vanishing contributions to the spectra in (\ref{mnp}) and (\ref{mnl}).

Moreover, we have applied the theoretical KK mode decompositions presented in sections 2 and 3, to the effective scales 
of couplings in the four-brane and five-brane cases respectively, and we then formulate the reasonable hierarchy between 
the Planck and electroweak scales. Namely, we have concluded that, for the given $M$ in the six and seven dimensions, 
we obtain 
\beq
M_{\rm eff}=Me^{-kr_{c}\pi}
\label{meff}
\eeq
as shown in (\ref{leffective6}) and (\ref{leffective7}). 

Finally, in phenomenological aspects, 
these KK modes in the extended RS scenario could be hopefully detected in the 
LHC detector, by exploiting the effective mass spectra (\ref{mnp11}) and (\ref{mnl10}) for ground states. These results are 
the same as (\ref{meff}) which is predicted by using the different paradigm as shown above. 
We also hope that the LHC could obtain the signals to indicate the effective mass spectra (\ref{mnp}) 
and (\ref{mnl}) for the excited states for the four- and five-branes, respectively.

In fact, the goal for early LHC running is around 1 $fb^{-1}$ at $\sqrt{s}=7~TeV$. 
Exploiting the statistical algorithms around in these ranges, they found that broad resonances to distinguish them from both 
background and contact interactions~\cite{kelly11}. Assuming that the Standard Model fields are supposed 
to reside on the $TeV$ brane, they concluded that the LHC could prove the full parameter space if the electroweak scale on the 
three brane is less than $10~TeV$~\cite{hewett01}. Moreover, including the Standard Model fermions in all six dimensions, they 
also considered the LHC phenomenology of the new gauge KK states which depend on the five dimensional fermions~\cite{rizzo08}.
On the other hand, the signals of LHC KK excitations of the electroweak gauge bosons have been investigated numerically 
in the paradigm with the Standard Model gauge and fermion fields propagating in a warped extra dimension~\cite{agashe07}. 
Similarly, they also predicted the production of KK gluons at LHC, which is suppressed by small couplings to 
the proton's constituents~\cite{agashe08}. The author would like to thank W.T. Kim, J. Lee, T.H. Lee, P. Oh and Y.J. Park for helpful discussions.\\

\end{document}